\documentclass[12pt]{article}
\usepackage[utf8]{inputenc}
\usepackage{textcomp}
\usepackage{amsmath}
\usepackage{amssymb}
\usepackage{graphicx}
\usepackage{braket}
\usepackage{url} 
\usepackage{comment} 
\usepackage{subfigure}
\usepackage[makeroom]{cancel}
\usepackage{setspace} 
\usepackage{hyperref}
\usepackage{mathptmx}
 
\usepackage{authblk}

\usepackage[legalpaper, portrait, margin=1in]{geometry}

  \usepackage[
    backend=biber,
    style=ieee,
  ]{biblatex}

\begin{document}

\pagenumbering{roman}


\title{\huge \vspace{-2.0cm} 
On measurement, superdeterminism, free will, and contextuality}


\author[1]{Mordecai Waegell}


\affil[1]{Institute for Quantum Studies, Chapman University, Orange, CA, USA}

\date{\today}

\maketitle

\begin{spacing}{1.2}
    \begin{abstract}
Superdeterminism has received recent attention as a possible path toward a locally causal explanation of the entanglement correlations that appear in experimental tests of Bell's theorem.  While the term `superdeterminism' was coined by Bell to refer to restrictions on the free will of experimenters, it was not rigorously defined until recently. It has now been defined as a property of any physical theory that produces systematic violations of statistical independence.  Here we focus on formalizing the requirements that being nonsuperdeterministic places on a physical theory, and setting a standard that must be met before we can conclude that a given theory is not superdeterministic.  We begin by carefully examining how a physical theory determines what outcomes we observe when performing measurements in terms of ontic states and response functions, and how this differs between superdeterministic and nonsuperdeterministic theories, in terms of the behavior of the types of vetted random sampling procedures that we use in experiments.  The core result is that individual samples and measurement outcomes must be representative of the observed distributions, which is explained in detail.  This also has a bearing on how measurement settings are chosen by agents, whether freely or randomly, and we argue that this standard ultimately defines what freedom/independence actually mean.  We then discuss contextuality, and show that in most cases, superdeterminism is contextual.  Finally, we discuss how different physical theories, with different notions of ontic states and response functions, can give rise to the same empirical data, and how the same operational contextuality may appear in different forms.
\end{abstract}
\end{spacing}

    {\bf Keywords:} Superdeterminism, statistical independence, measurement independence, Bell's theorem, locality, free will, retrocausality.

\pagebreak

\begin{spacing}{1.8}
\tableofcontents
\end{spacing}
\newpage

\section{Introduction}\label{intro}
\pagenumbering{arabic}
\setcounter{page}{1}

What does it mean for us to perform scientific measurements?  What are we actually revealing by measuring?  For this article, we will adopt the position that one can only begin to answer these questions when one is referring to a specific explanatory physical theory $T$.  The only restrictions we place on the types of physical theories we will consider are that they be self-consistent and that they explain the relevant empirical data.  This may include superdeterministic theories, theories where causality does not follow temporal order, theories which lack causality altogether, and many others.  For most standard types of theories, both the laws and the initial/boundary conditions must be specified as part of $T$, assuming both are needed for that theory to produce the empirical data. 

It seems beyond argument that there must be some theory $T$ that is ontologically true, and whose mechanisms explain the empirical data we observe, even if we humans may never know which $T$ that is.  However, if a human has a theory to propose, then we also require that all of the mechanisms of $T$ that are needed to generate the relevant empirical data are fully defined, so that we can see clearly how they work, and that they are self-consistent - otherwise there is no reason to take that $T$ seriously (this approach was first developed in \cite{waegell2023generative}).\footnote{For example, relational quantum mechanics (RQM) does not meet this standard, so it cannot (yet) be seriously considered.  RQM aspires to a particular set of concepts, but has yet to provide a coherent local theory that explains agreement between multiple observers (\cite{rovelli1996relational, adlam2023information, di2025relative} ).   Ideas like QBism (\cite{fuchs2014introduction, fuchs2023qbism}) are not even proposed as physical theories, so they also fall outside the scope of this work.} Once we can see the mechanisms of $T$, it should be straightforward to assess whether the theory obeys constraints like \textit{local causality} or \textit{no superdeterminism} pertinent to Bell's theorem (\cite{Bell2, bell1990nouvelle}) - provided those constraints are also well defined.  One of the main purposes of this article is to clearly define the constraint that \textit{no superdeterminism} places on the mechanisms of a theory.  This article is intended as a companion and  follow-on to a recent work which establishes a rigorous definition of superdeterminism itself \cite{waegell2025statistical}: \textit{Superdeterministic theories are those in which there are situations where it is impossible to use a good random sampling procedure to obtain a representative sample.}  That article also gives a thorough literature review of the topic, which is not repeated here for the sake of brevity.

To begin, we need to discuss how measurements are understood in physical theories.  Borrowing from the ontological models framework \cite{harrigan2010einstein}, in general, measurement works in two distinct steps: First, it is somehow determined which single specific \textit{ontic state} $\lambda$ one is performing the measurement upon, and second, the measurement device interacts in some way with that $\lambda$ and an outcome $k$ is encoded into the measurement device.

In many physical theories we ascribe $\lambda$ to individual elements of an ensemble, or equivalently, to individual runs of an experiment.  For example, we might ascribe an individual $\lambda$ to each individual electron in an ensemble, or each individual person in a clinical drug trial, and then by preparing/selecting a single electron/person, we have prepared/selected a single $\lambda$.

However, some physical theories contain spatially and/or temporally extended nonseparable objects.  For example, quantum mechanics contains spatially nonseparable entangled states where the ontic state $\lambda$ that explains the outcomes measurements on the joint system or its subsystems cannot be reduced to separate descriptions of the subsystems that make no reference to one another.  When we measure one of the entangled subsystems alone, we are measuring this collective $\lambda$, and not an individual state ascribed only to this subsystem, and it is not necessarily the case that each electron in an ensemble has its own distinct state.  Thus, in the case of an experiment with many components, we often associate a single $\lambda$ with each run of the entire experiment, which describes all of the relevant systems at once.

For example, when considering Bell's theorem, $\lambda$ for a single run is usually understood as the ontic state describing all of the prepared systems in the experiment, without presuming they are separable at the start.  Bell's factorization condition arguably amounts to the requirement of spatial separability, such that $\lambda$ must decompose into set of local hidden variables ascribed to the individual particles.

We could also imagine an extreme holistic theory wherein all objects in space and time belong to a single nonseparable entity with a single $\lambda$ which determines the response to all measurements.  In such a theory, our attempts to isolate individual elements/runs with separate $\lambda$s always fail, since every element in an ensemble shares the one collective $\lambda$.

As such, a human-provided, spatially or temporally nonseparable theory must be very specific about what $\lambda$s one is measuring when selecting out individual elements of an experimental ensemble, and how they explain the empirical data.

We now focus on the first step of the measurement process; how one selects the individual $\lambda$ from within an ensemble.  In scientific experiments, we often use \textit{random sampling procedures} in an attempt to obtain \textit{representative samples} from the ensemble we wish to learn about.  A \textit{good} random sampling procedure $R$ is one we have applied many times to systems with an ontic property in $\lambda$ that we can see, and for which we have never observed any unexpected correlation between $\lambda$ and the random selections.  The assumption of \textit{statistical independence} (SI) is that when we apply a good random sampling procedure to select a sub-ensemble from within an ensemble where $\lambda$ has ontic properties we \textit{cannot} see (hidden variables), there is still no correlation.  Under this assumption, the relative frequency $\rho(\lambda)$ of a given $\lambda$ is approximately the same in the ensemble as in the sub-ensemble (i.e, $\rho(\lambda | R) = \rho(\lambda)$), which means the sub-ensemble is a representative sample of the ensemble from which it was drawn. 

However, there are superdeterministic theories \cite{brans1988bell, t2016cellular, donadi2020superdeterministic, hance2022supermeasured, garza2025quantum}, for which statistical independence is systematically violated, wherein all good random sampling procedures may sometimes
fail to produce representative samples - even if we attempt to use our own free choice to make the selection.  We tend to imagine that our own choices, like how we select a bingo ball from a hopper, or a card from a deck, are separate from the laws of physics - that we are somehow independent free agents and that the laws simply define the theater in which we operate.  Even if we accept that our choices are governed by physical laws, we tend to imagine that the details governing the choice are independent of the details governing the cards in the deck.  But in superdeterministic theories, these details may not be independent, and there may be conspiratorial-seeming correlations between the free or random selections and the $\lambda$s that are selected.  

To illustrate, consider a toy theory where whenever we attempt to take a deck of standard playing cards and randomly divide them into two equally sized facedown piles, we then find that all of the cards in the first pile are red, and all of the cards in the second pile are black.  In this toy theory, the correlation between $\lambda$ and the choices can be directly observed after the fact, which would actually cause us to reject whatever random sampling procedure we had used to divide the cards.  This does not make the toy theory superdeterministic, because our empirical data would lead us to expect this correlation.  In a superdeterministic theory, the same type of correlation may occur with hidden aspects of $\lambda$, even when we have no empirical reason to reject the free/random sampling procedure we are using - i.e., even when we use a good random sampling procedure. \footnote{If a theory makes good random sampling procedures entirely impossible in some scenario, meaning we always observe correlations, and never get representative samples, no matter what we try, then this also satisfies the definition of superdeterminism.  In the toy theory, if every blind sampling procedure we could try results in all red cards in the first pile, and all black in the second, then we cannot obtain representative samples, and the toy theory is superdeterministic. This last case would only be relevant if future technology allowed us to observe all of these correlations, so that we could clearly see that our choices were not free/random.}

The key point to take away here is that the best we can do as agents/experimenters is to perform a good random sampling procedure $R$ (presuming the theory allows them), but it is the theory itself which dictates which specific $\lambda$ we get each time, and the frequencies $\rho(\lambda |R)$ with which they end up in our sample.  In nonsuperdeterministic theories, this $\rho(\lambda |R)$ is a representative sample of the larger ensemble from which it was drawn, and in superdeterministic theories it may not be, and this is entirely determined by the theory, regardless of how good we think our free/random sampling procedure is. \footnote{Strictly speaking, it may be the conjunction of the laws in a physical theory and a fluke postulate that result in superdeterminism \cite{waegell2025statistical}, but for brevity here, we will always attribute this collectively to the physical theory $T$.}

As such, we think it is helpful in what follows to think of $\rho(\lambda |R)$ as a kind of response function which describes how the theory responds to the implementation of a random sampling procedure by providing a specific ontic state $\lambda$.  This may seem counterintuitive, since it may be hard to see how the theory is responding by providing a specific bingo ball when I reach into the hopper, given that I was just choosing from among things that were already there, with a well-defined relative frequencies $\rho(\lambda)$ - but it was the theory $T$ that determined where those bingo balls were in the first place, and whether or not my choice is correlated with the ontic states of the bingo balls.  This is the abstract sense in which the theory $T$ \textit{responds} when we implement good random sampling procedures.  Thus, whether superdeterministic or not, it is always the theory that provides an ontic state $\lambda$ in response to a random sampling procedure $R$, using the \textit{random sampling response function}, $\rho(\lambda |R)$.

Note that because we are referring to finite experimental ensembles, we are always thinking of $\rho$ as a relative frequency rather than a probability.  Furthermore, in deterministic theories, where nontrivial probabilities play no physical role, only relative frequencies are meaningful.  Some physical theories may contain well-defined notions of infinity, while others may not.  Some theories may make genuine use of probabilities and stochastic processes, while others may not.  Either way, real experimental ensembles are finite.  For very large ensembles, results from probability theory like Bayes' theorem may be approximately applicable, but must be used with caution.  

We now move to the second step of the measurement process; once we have selected out a $\lambda$, that $\lambda$ determines the outcome(s) $k$ for any operational measurement procedure $M$ that we could perform in the lab.  In deterministic theories, $\lambda$ specifies definite outcomes $k$ for each possible $M$, whereas indeterministic theories may instead specify a nontrivial probability distribution $\xi(k|M,\lambda)$ over different outcomes, sometimes called a \textit{response function} \cite{spekkens2005}.  To avoid ambiguity, we will call this a \textit{measurement response function}.

These two steps and their two response functions combine to give us the long-run empirical probability/frequency of getting outcome $k$ for measurement procedure $M$ on a randomly sampled element from an ensemble containing each $\lambda$ with relative frequency $\rho(\lambda)$,
\begin{equation}
     \textrm{Pr}(k|M,R) = \sum_{\lambda \in \Lambda} \xi(k|M, \lambda)\rho(\lambda|R).  \label{PkM}
\end{equation}

In a nonsuperdeterministic theory we have $\rho(\lambda |R ) = \rho(\lambda)$, and the expression for the empirical statistics simplify to,
\begin{equation}
     \textrm{Pr}(k|M,R) = \sum_{\lambda \in \Lambda} \xi(k|M, \lambda)\rho(\lambda).  \label{PkM_NS}
\end{equation}
for any good random sampling procedure $R$.  In superdeterministic theories, the results of $\rho(\lambda|R)$ may be correlated with $M$ in some conspiratorial way.

It is worth noting this expressions specifies the probabilities for indeterministic theories, but a fluke postulate indicating the we had obtained atypical results from $\xi(k|M, \lambda)$ could still result in a superdeterministic theory.

Given that it is the physical theory which determines both response functions $\rho(\lambda|R)$ and $\xi(k|M, \lambda)$, we can also think of the theory as having a single \textit{generalized response function} $\mathcal{R}(k|M,R)$ that produces the empirical statistics $\textrm{Pr}(k|M,R)$ given random selection procedure $R$ and measurement procedure $M$ - regardless of any details about $\lambda$s.

The notion of typicality of a distribution plays a significant role in our arguments.  A typical sequence is one with no special patterns or properties, which thus produces approximately the average result.  For example, if I flip a million fair coins in a row, the vast majority of possible sequences have approximately equal numbers of heads and tails, and no special pattern in how they are arranged, and these are the typical sequences.  Cases like all heads, or a perfect repeating pattern heads-tails-heads-tails... are atypical, which also means they are sequences of lower complexity.  Importantly, if we consider an indeterministic theory with stochastic fair coins, it is the typical sequences which are truly representative of the 50\% stochastic probability used to generate the sequence.

\section{Preparation procedures}

So far, we have not given any consideration to where the ensembles from which we wish to draw representative sub-ensembles themselves originate.  

For this, we must now discuss operational \textit{preparation procedures} which we can perform in order to obtain an ensemble.  We define a preparation procedure $p$ as any procedure in the set $P$ that measures and verifies that the system has the properties in a list $L_P$, e.g., if we want to do a drug trial with children between ages 7 and 12, we measure and verify that the ages of the children fall within this range before we include them in the trial - and there may be different ways to obtain this information, e.g., consulting birth records, measuring some present property that is a good indicator of age, or running a stopwatch since their birth.  Strictly speaking, we will consider any system which has all of the properties in $L_P$, whether or not we have observed them, to have been prepared by a procedure in $P$.  The elements in the prepared ensemble are left with an unknown distribution over any properties not in the list $L_P$, characterized as $\rho(\lambda |p)$, since the preparation $p$ is blind to those properties.

Since the goal in preparing an ensemble for an experiment is to create a representative sample of all systems with the properties in list $L_P$, experimentalists always use good random selection procedures to select the elements to check for properties $p$.  Thus, a good preparation procedure $p$ always includes a good random selection procedure, and a random selection procedure by itself is simply a preparation $p$ with $L_P$ empty. Thus, for a particular experimental ensemble preparation procedure $p \in P$, statistical independence is the assumption that $\rho(\lambda | p) \approx \rho(\lambda|P)$, i.e., we assume our randomly selected sample of children ages 7 to 12 is representative of all children in that age range. As an illustration, this also means our good random selection process must have sampled children from all over the world, since if we only sample children in, say, Germany, then the list $L_P$ includes, `in Germany' as well as `ages 7-12,' and we would only expect the ensemble to be representative of children in Germany.  In that case, we might still develop a theory that extrapolates about all other children in the world, but this is a separate issue.  There are other properties in $L_P$ for the children, like, `on Earth,' and `human,' which naturally restrict what we expect our sample to be representative of - e.g., human children on Earth between ages 7 and 12.

The physical theory also determines which systems in our sample will have the properties in $L_P$, and thus it determines a preparation response function $\rho(\lambda | p)$, just as before.   If we are working with a prepared experimental ensemble with non-empty $L_P$, then $\rho(\lambda | p)$ replaces $\rho(\lambda|R)$ and $\rho(\lambda | P)$ replaces $\rho(\lambda)$ in all of the above expressions.  Thus, the probability of obtaining outcome $k$ given measurement procedure $M$ and preparation procedure $p$ is given by,
\begin{equation}
     \textrm{Pr}(k|M,p) = \sum_{\lambda \in \Lambda_p} \xi(k|M, \lambda)\rho(\lambda|p),  \label{PkMp}
\end{equation}
and a nonsuperdeterministic theory produces typical results from,
\begin{equation}
     \textrm{Pr}(k|M,p) = \sum_{\lambda \in \Lambda_P} \xi(k|M, \lambda)\rho(\lambda|P),  \label{PkMP}
\end{equation}

\section{The standard for nonsuperdeterministic theories}




  The fact that all nonsuperdeterministic theories must use generalized response functions $\mathcal{R}(k|M,p)$ that are representative of  Eq. \ref{PkMP} is a strong constraint on those theories, so we take a moment here to unpack it.    First, our past observations establish the \textit{empirically plausible} statistics $\textrm{Pr}(k|M,p)$ of outcomes $k$ for each pairing of a measurement $M$ and a preparation $p$, which tells us about the form of $\mathcal{R}(k|M,p)$ in the theory we are working with -- complete with statistical flukes, fine-tuned initial conditions, or any nomic restrictions.  For example, when we prepare a Bell state, observations that obey the entanglement correlations (to within noise tolerances) are empirically plausible, and observations that violate them are not, since this is what we have always observed.
  
  Now, If we use a good random selection procedure to select a single system from a prepared ensemble, it must be the case that the frequency of that single element having ontic state $\lambda$ is $\rho(\lambda|p)$, and this must be true for any single element we select from the ensemble: every randomly selected element has the possibility of being in every ontic state $\lambda$ with nonzero frequency in $\rho(\lambda|p)$.  If we consider randomly subjecting each element in the ensemble to a different measurement $M$, it then follows that the response $\xi(k|M,\lambda)$ must be well-defined and empirically plausible for every pairing of an $M$ with a $\lambda$ with nonzero frequency in $\rho(\lambda|p)$, because for every $M$ we are going to perform on a randomly selected element, that element has the possibility of having those $\lambda$.  
  Another way to say this is that in a nonsuperdeterministic theory, the measurement response function $\xi(k|M,\lambda)$ must define an empirically plausible outcome for all \textit{counterfactual} measurements $M$, in addition to the one measurement that actually occurs.

  We can thus think of generalized response functions $\mathcal{R}(k|M,p)$ as tables conditioned on $p$ that list an empirically plausible response for every setting $M$.  This response may not be a specific outcome, but may instead trigger an outcome to be selected from a particular probability distribution.  Indeed, if what is empirically observed in the long-run statistics for a given pairing of preparation $p$ and measurement $M$ is not a single definite outcome, but instead a nontrivial frequency distribution $\textrm{Pr}(k|M,p)$ (like the Born rule distribution in quantum experiments), then it is this distribution that we deem empirically plausible for that $M$ and $p$, and thus every single randomly selected element must be representative of this distribution.  Note that empirical samples are always finite and subject to experimental limitations, so what is actually empirically plausible is a relative frequency, and never a true probability.  However, a physical theory $T$ may specify an actual probability distribution (like the Born rule), or a relative frequency for an infinite ensemble, which is essentially equivalent, and then we can check to see if $\mathcal{R}(k|M,P)$ for individual elements is representative of a typical distribution of the specified type.  
  
  It is not sufficient that a large sub-ensemble reproduce this distribution - it must be the case that each individually selected element is representative of this entire distribution, whether or not other elements are ever examined.  Note that this does not automatically require the theory to be indeterministic, since even in a deterministic theory, a good random sampling procedure can still produce representative samples.
  
  This last requirement is very subtle.  A single element has a single $\lambda$, so clearly that cannot be representative of an entire distribution $\rho(\lambda | P)$, but that is not the idea.  When we apply a good preparation procedure with a good random selection procedure to select a single element, then if the physical theory is not superdeterministic, that single element will have ontic state $\lambda$ with frequency $\rho(\lambda |
  P)$, and this is the sense in which a single randomly selected element is representative of the ensemble frequency - it is a question of how the theory treats the random selection procedure.  As experimenters, all we can do is attempt to select a single element at random, but it is always the theory's preparation response function that determines the specific $\lambda$ we obtain for that element.
  
  If each individual selection is not independently representative of the distribution $\rho(\lambda|P)$, and no other distribution, this represents a superdeterministic correlation between the selections and the ontic states.  
  
  As a fanciful example, suppose a magical goblin is responsible for choosing $\lambda$ each time you prepare a Bell state for a Bell experiment.  In a given run, the goblin freely chooses a specific $\lambda_0$ and then controls the settings chosen by the experimenters so that the chosen $\lambda_0$ produces empirically plausible outcomes for those settings, even if it has empirically implausible outcomes defined for all other settings.  By controlling every run and choosing different $\lambda_0$ and settings each time, the goblin can reproduce the empirical statistics $\textrm{Pr}(k|M,p)$ and also make it so that each (joint) measurement setting is chosen with roughly equal frequency, and no apparent pattern, so that they appear free/random.  The hidden statistics $\rho(\lambda|P)$ could even be satisfied for long runs, while individually $\rho(\lambda_0|M_0,p) = 1 \neq \rho(\lambda|P)$. Nevertheless, the facts that the freely chosen $\lambda_0$ are not representative of $\rho(\lambda|P)$, and only certain $M_0$ are allowed for each $\lambda_0$, enable the goblin to reproduce the Bell statistics without violating local causality.  As in this case, if the mechanisms of the physical theory (e.g. goblins) are representative of any information more specific about $\lambda$ than $\rho(\lambda|P)$ on a single run, then those mechanisms will generally produce superdeterministic correlations.

  This establishes a standard that must be shown for any physical theory before we can be sure it is not superdeterministic: 
  \begin{quote}
      Both the ensemble of empirical data, and the generalized response function $\mathcal{R}(k|M,p)$ used to produce each individual outcome, must be representative of a typical distribution given preparation $P$ and the relative frequency $\rho(\lambda|P)$, and measurement $M$ and the measurement response function $\xi(k|M,\lambda)$ specified by the theory.
  \end{quote}

Meeting the standard we have established here for nonsuperdeterminism rules out any kind of hidden or conspiratorial correlation between $\lambda$ and $M$, and we think this is what it means for these two things to be genuinely free of each other and physically independent.  This definition allows for two things to be deemed independent even in a deterministic theory where they share common causes.

This standard raises a particular challenge for Bohmian mechanics \cite{bohm1952suggested}, Everett many-worlds \cite{everett1957relative,wallace2010quantum}, and any other theory that contains nonseparable entangled wavefunctions in its ontology.

These theories must explicitly show how separate effective ontic states $\lambda'$ and settings choices $M'$ emerge from within the generally nonseparable ontology, or there is no hope of establishing that they are independent.

This may be seen as a mark in favor of separable theories of quantum mechanics \cite{deutsch2000information,bedard2021abc, kuypers2021everettian, waegell2023local, kuypers2026restoring}. 

However, even in that case, entanglement can create exactly the types of conspiratorial correlations we are concerned with (e.g., $|
\lambda_1\rangle|M_1\rangle + |
\lambda_2\rangle|M_2\rangle $), meaning the generalized response functions in quantum theories do in fact allow for superdeterminism.  Thus, for two systems to be proved independent in such theories, it must also be shown that their effective descriptions are entirely separate (e.g., a product state).

  \section{Agents, free will, and random choices}

  Part of specifying a satisfactory physical theory $T$ is explaining how agents/experimenters are positioned within the physical theory, or, if they are not within $T$, then how they interact with $T$ in order to observe empirical data and make interventions upon the physical world.  Given that some of the superdeterministic theories we have been discussing place constraints on which measurement settings an agent is physically able to choose, it seems they must somehow exist within such theories.  What can we say about agents in nonsuperdeterministic theories, where the preparation response function is statistically independent of the measurement settings chosen by a good random selection procedure, i.e., $\rho(\lambda|M) = \rho(\lambda)$?  If we approximate the relative frequency of ontic state $\lambda$ as a probability, and also treat our free choices or good random selections as random variables with probabilities $\textrm{Pr}(M)$, then Bayes' theorem gives us
  \begin{equation}
      \textrm{Pr}(M|\lambda) = \rho(\lambda|M)\frac{\textrm{Pr}(M)}{\rho(\lambda)} = \textrm{Pr}(M),
  \end{equation}
where the last step uses the statistical independence formula.

Thus, in nonsuperdeterministic theories, we also have a symmetric constraint on the relative frequency and/or probability of choosing each measurement setting $M$: The relative frequency of choosing each setting given a particular ontic state $\lambda$ must be representative of the overall frequency of choosing that setting.  This again establishes that the theory must specify an empirically plausible response for every pairing of an $M$ with a $\lambda$.  And again, we can think of the agent's choice as the output of an \textit{agent response function} $\mathcal{A}(M)$.  For an agent making free choices, the empirically plausible experience seems to be the freedom to make any choice on any particular occasion, perhaps with some frequency distribution depending on the circumstances, so in a nonsuperdeterministic theory, each individual output from $\mathcal{A}(M)$ must be representative of that plausible distribution.

This applies to any good random selection procedure just as it does to free choices, and establishes that each flip of a good coin must be representative of a 50/50 distribution of heads and tails outcomes, assuming the theory $T$ makes this the typical distribution for that coin.

Now, in superdeterministic theories, we may have $\textrm{Pr}(M) \neq \textrm{Pr}(M |\lambda)$, and indeed we could have a case like $\textrm{Pr}(M) =1/2$, $ \textrm{Pr}(M |\lambda) = 0$, meaning that an empirically plausible choice $M$ is physically impossible for ontic state $\lambda$. \footnote{See, for example, the counterfactual restrictions of \cite{hance2024counterfactual}.}  That is, the agent believes they could have chosen $M$, but they actually could not have, given $\lambda$.  This is a direct constraint on the agent's ability to make free choices, and thus there is a clear sense in which superdeterminism constrains the free will of agents within such theories.

Likewise, in a superdeterministic theory, a random coin may have no physical chance of coming up heads on a given toss, even if the empirically plausible frequency is 50/50.

\section{Superdeterminism is (usually) preparation contextual}

If two different procedures $p$ and $p'$ produce the same measurement statistics $\textrm{Pr}(k|M,p) = \textrm{Pr}(k|M,p')$ for all measurements $M$ then we say that $p$ and $p'$ are \textit{operationally equivalent preparations} \cite{spekkens2005}, denoted $p\simeq p'$.  Operationally equivalent preparation procedures need not involve the same list of observed properties $L_P$.  For example, we can prepare a maximally mixed state of a qubit by $p$, using a machine that randomly outputs up or down $\hat{\sigma}_x$ eigenstates with equal frequency, or by $p'$, using a  machine that randomly outputs up or down $\hat{\sigma}_y$ eigenstates with equal frequency.  Even though the the empirical statistics are the same for $p$ and $p'$, the lists of observations we have made about them (e.g., which machine we used to prepare them) are different.

On the other hand, it seems that all procedures which do verify the same list of properties $L_P$ (all $p$ in $P$) should also be operationally equivalent, but perhaps counterexamples exist that we are missing.

A theory is said to be \textit{preparation noncontextual} if and only if any two operationally equivalent preparation procedures result in the same frequency of ontic states, i.e., iff $\rho(\lambda|p) = \rho(\lambda|p')$ whenever $p\simeq p'$.

Now, if we consider a specific pair of preparation procedures $p$ and $p'$ belonging to a set $P$ which verifies the properties in list $L_P$, and it is also true that $p \simeq p'$, then a violation of statistical independence of the form $\rho(\lambda|p) \neq \rho(\lambda|p')$ is also a violation of preparation noncontextuality.  There is some subtlety to unpack here.  In general, SI requires $\rho(\lambda|p) = \rho(\lambda|p')  = \rho(\lambda|P)$, but strictly speaking, it would still be violated if $\rho(\lambda|p) = \rho(\lambda|p')  \neq \rho(\lambda|P)$, which would not constitute a violation of preparation noncontextuality.  For example, if we randomly draw two sub-ensembles from a large ensemble, they may both fail to represent the larger ensemble in the exact same way, such that they do represent one another.  While it seems possible for a physical theory to work like this, it seems that a much broader class of theories would have $p$ and $p'$ fail to represent $P$ in arbitrarily different ways, and in all such theories, violations of SI are also violations of preparation contextuality.

A less formal version of this observation was made in \cite{allori2024hidden}.

\section{Measurement procedures and measurement contextuality}

Now, we should emphasize a subtle issue here about measurement contexts.  The operational measurement procedure $M$ may generally involve the measurement of more than one observable property $O$ of the system at once, and $\lambda$ specifies how an outcome $k$ is determined for each distinct property.  In general, many different measurement procedures can measure the same observable property $O$.  If two different measurement procedures $M$ and $M'$ both implement measurements of $O$, which has possible outcomes $k$, and always give the same empirical statistics $\textrm{Pr}(k|M) = \textrm{Pr}(k|M')$ for identically prepared systems, then we say that the pairs $[k,M]$ and $[k,M']$ are \textit{operationally equivalent} ways of measuring $O$, denoted $[k,M] \simeq [k,M']$ \cite{spekkens2005}. We call the set of observable properties $\{O_i\}_M$ that are all measured together by procedure $M$ the \textit{measurement context} associated with $M$.  In \textit{measurement noncontextual} physical theories, the probability distribution $\xi(k|M,\lambda)$ over the outcomes of a particular observable cannot depend on which procedure from an operationally equivalent set is implementing the measurement of $O$, which is to say, if $[k,M] \simeq [k,M']$, then it must be true that $\xi(k|M,\lambda) = \xi(k|M',\lambda)$.  In other words, the measurement response function that determines $k$ cannot depend on the context in which $O$ is measured.  The notion of measurement noncontextuality comes from our experience of the macroscopic world, where for example, the color seen on one face of a cube does not depend on which other faces are being seen at the same time (leaving aside special optical trickery like holograms).  Under certain seemingly common-sense assumptions, including no superdeterminism, quantum mechanics is found to be measurement contextual, in contradiction of our classical experience.  Specifically, there is no consistent way to predefine a single (noncontextual) value $k$ for the outcomes of all observables $O$.

\section{The slippery boundary between $\xi$ and $\rho$.} 

There are many different ways to construct a theory so that the same empirical frequencies $\textrm{Pr}(k|M,p)$ are observed.  At one extreme are theories with only a single nonseparable ontic state $\lambda$ that determines the outcomes of all measurements on all elements of any experimental ensemble, so $\rho(\lambda)=1$ (and $\rho(\lambda')=0$ for all $\lambda'\neq\lambda$), and any uncertainty about the outcomes originates in the response function $\xi(k|M, \lambda)$.  At the opposite extreme are theories where there are distinct $\lambda$s for each element in the ensemble, while for each $\lambda$, the response function is trivial (e.g., there is a single definite outcome $k$, so that $\xi(k|M, \lambda) = 1$ and $\xi(k'|M, \lambda) = 0$ for all $k'\neq k$), and all of the uncertainty originates in the selection from $\rho(\lambda)$.  Bohmian mechanics is an example of the latter type of theory, since once the Bohmian hidden variable $\lambda$ is known, there is no remaining uncertainty about the outcomes of measurements - although the ensemble $\Lambda$ from which this hidden variable is drawn is metaphysical, and not something we can sample from experimentally. In fact, all details about actual experimental ensembles in Bohmian mechanics do follow from the single $\lambda$, so in this sense Bohmian mechanics is also like the first extreme, except that $\lambda$ may be separable.

This slipperiness extends to the issue of measurement and preparation contextuality.  Consider a theory that is measurement contextual, meaning that a given $\lambda$ may specify different outcome probabilities for measurement of the observable property $O$, depending on whether the measurement procedure $M$ or $M'$ implements the measurement of $O$.  The same empirical statistics $P(k|M,p)$ and generalized response function $\mathcal{R}(k|M,p)$ can be produced by a different theory that violates preparation contextuality instead of measurement contextuality.  In this theory, the distribution produced by the preparation procedures must depend explicitly on what measurement setting will be applied later, $\rho(\lambda|p,M)$, which is a superdeterministic (or perhaps retrocausal) correlation, but then each $\lambda$ may have a noncontextual measurement response function.  The outcomes for $O$ will still differ between $M$ and $M'$ in the same way, but now it is because we get different $\lambda$s for each measurement context, rather than getting different outcomes from the same $\lambda$.  Thus, a special class of superdeterministic preparation contextual theories can reproduce all of the empirical properties of nonsuperdeterministic measurement contextual theories.  Given that the prepared ensemble depends explicitly on the future measurement setting $M$, one might still call this `measurement contextuality,' even if this is not the usual mechanism.  

The point here is that the empirical statistics are related to the generalized response functions $\mathcal{R}(k|M,p)$, which is somewhat oblivious to how the theory deals with ontic states $\lambda$ and thus focusing on the properties of $\rho(\lambda|p)$ and $\xi(k|M,\lambda)$ may be an unnecessary distraction in some cases.  If $\mathcal{R}(k|M,p) \neq \mathcal{R}(k|M',p)$ when $[k,M] \simeq [k,M']$ then the theory appears to be measurement contextual, regardless of the details.

\section{Conclusion}

We have unpacked how general physical theories produce responses to measurements, introducing response functions for random sampling and preparation procedures, and incorporating them with the standard measurement response functions in order to define a theory's generalized response functions.  With this, we have given a formal standard that theories must satisfy before they can be deemed nonsuperdeterministic.  The idea here is to put the onus on the proponents of a given theory to explain all of its mechanisms in a consistent and concise way, and thus to show whether or not it meets the standard of nonsuperdeterminism.   From here, we also set a standard for choices by agents to be truly random or free.  Finally, the connections between superdeterminism and contextuality have been formalized in detail.\newline

\noindent \textbf{Acknowledgments:}  This project/publication was made possible through the support of Grant 63209 from the John Templeton Foundation. The opinions expressed in this publication are those of the authors and do not necessarily reflect the views of the John Templeton Foundation.\newline

\noindent\textbf{Conflicts of Interest:} The corresponding author states that there is no conflict of interest. \newline

\noindent\textbf{Data Availability:} There is no data associated with this manuscript.


\printbibliography
\end{document}